\documentclass[prd,aps]{revtex4}

\usepackage{graphicx,subfigure}

\newcommand{\be}{\begin{equation}}
\newcommand{\ee}{\end{equation}}
\newcommand{\bea}{\begin{eqnarray}}
\newcommand{\eea}{\end{eqnarray}}

\begin{document}


\title{
CP violating dimuon charge asymmetry in general left-right models
}

\author{ Kang Young Lee }
\email{kylee14214@gmail.com}

\affiliation{
Division of Quantum Phases \& Devices,
School of Physics, Konkuk University, Seoul 143-701, Korea
}

\author{ Soo-hyeon Nam }
\email{glvnsh@gmail.com}

\affiliation{
Supercomputing Center, KISTI, Daejeon 305-806, Korea
}

\date{\today}

\begin{abstract}

The recently measured charge asymmetry of like-sign dimuon events
by the D0 collaboration at Tevatron shows the 3.9 $\sigma$ deviation
from the standard model prediction.
In order to solve this mismatch,
we investigate the right-handed current contributions
to $B_s-\bar{B}_s$ and $B_d-\bar{B}_d$ mixings
which are the major source of the like-sign dimuon events
in $b \bar{b}$ production in general left-right models
without imposing manifest or pseudo-manifest left-right symmetry. 
We find the allowed region of new physics parameters 
satisfying the current experimental data.

\end{abstract}

\maketitle

\section{Introduction}

Recently the D0 collaboration has measured a deviation
from the standard model (SM) prediction in the CP violating like-sign dimuon
charge asymmetry in semileptonic $b$ hadron decay
with the 9 fb$^{-1}$ integrated luminosity of $p \bar{p}$ data
at Tevatron \cite{d0new}:
\be \label{asym_exp}
A_{sl}^b = -0.00787 \pm 0.00172~({\rm stat.}) \pm 0.00093~({\rm syst.}) .
\ee

The like-sign dimuon events 
comes from direct semileptonic decays
of one of $b$ hadrons and a semileptonic decay of the other $b$ hadron
following the $B^0-\bar{B}^0$ oscillation in $b \bar{b}$ pair production at Tevatron,
defined by
\be
A_{sl}^b \equiv \frac{ \Gamma(b \bar{b} \to \mu^+ \mu^+ X)
                       - \Gamma(b \bar{b} \to \mu^- \mu^- X)}
                       { \Gamma(b \bar{b} \to \mu^+ \mu^+ X)
                       + \Gamma(b \bar{b} \to \mu^- \mu^- X)}.
\ee
At Tevatron experiment,
both decays of $B_d$ and $B_s$ mesons contribute to the asymmetry.
If we define the charge asymmetry of semileptonic decays of neutral $B^0_q$ mesons as 
\be
a_{sl}^q \equiv \frac{\Gamma(\bar{B}^0_q(t) \to \mu^+ X)
                     - \Gamma(B^0_q(t) \to \mu^- X)}
                     {\Gamma(\bar{B}^0_q(t) \to \mu^+ X)
                     + \Gamma(B^0_q(t) \to \mu^- X)},
\ee
the like-sign dimuon charge asymmetry can be expressed
in terms of $a_{sl}^q$ as \cite{grossman}
\be \label{Aslb}
A_{sl}^b = \frac{1}{f_d Z_d + f_s Z_s}
           \left( f_d Z_d a_{sl}^d + f_s Z_s a_{sl}^s \right),
\ee
assuming that $\Gamma(B_d^0 \to \mu^+ X)=\Gamma(B_s^0 \to \mu^+ X)$ to a very good approximation, 
where $f_q$ are the production fractions of $B_q$ mesons, and
$Z_q = 1/(1-y_q^2)-1/(1+x_q^2)$
with $y_q = \Delta \Gamma_q/(2 \Gamma_q)$, $x_q=\Delta M_q/\Gamma_q$.
These parameters are measured to be
$f_d = 0.402 \pm 0.013$,
$f_s = 0.112 \pm 0.013$,
$x_d = 0.771 \pm 0.007$,
$x_s = 26.3 \pm 0.4$,
and
$y_d = 0$,
$y_s = 0.052 \pm 0.016$
\cite{HFAG}.
With these values, Eq. (\ref{Aslb}) is rewritten by
\be
A_{sl}^b = (0.572 \pm 0.030) a_{sl}^d + (0.428 \pm 0.030) a_{sl}^s.
\ee

The non-zero dimuon asymmetry implies 
a difference between the $B^0 \leftrightarrow \bar{B}^0$ transitions 
and the CP violation in the $B$ system.
In the SM, the source of the CP violation
in the neutral $B^0_q$ system is the phase of
the Cabibbo-Kobayashi-Maskawa (CKM) matrix elements involved
in the box diagram.
Using the SM values for the semileptonic charge asymmetries 
$a^d_{sl}$ and $a^s_{sl}$ of $B^0_d$ and $B^0_s$ mesons,
repectively \cite{Lenz07}, 
the prediction of the dimuon asymmetry in the SM is given by
\be
A_{sl}^b = (-2.7^{+0.5}_{-0.6}) \times 10^{-4},
\label{Asl_SM}
\ee
which shows that the D0 measurement of Eq. (1) deviates
about 3.9 $\sigma$ from the SM prediction.
If the deviation is confirmed with other experiments,
it indicates the existence of the new physics beyond the SM.
Recently several works are devoted to explaning
the D0 dimuon asymmetry in the SM and beyond \cite{dimuon}.

As an alternative model solution to the mismatch 
between the measurement and the SM prediction of the dimuon charge asymmetry,
we consider the left-right model (LRM) 
based on the SU(2)$_L \times$ SU(2)$_R \times$ U(1) gauge symmetry
which is one of the attractive extensions of the SM \cite{LRM}.
The current measurement of the dimuon charge asymmetry 
can be explained in the LRM due to
the sizable right-handed current contributions to $B^0-\bar{B}^0$ mixing 
\cite{Nam02}.
The manifest left-right symmetry provides an natural answer 
to the origin of the parity violation.
Involving the triplet Higgs field $\Delta_{L,R}$ 
to break the additinal SU(2)$_R$ symmetry,
the lepton number violating Yukawa terms are introduced
and the see-saw mechanism for light neutrino masses is exploited in the LRM.
This model arises as an intermediate theory 
in the SO(10) grand unified theory (GUT).
In the LRM, the right-handed fermions transform as doublets 
under SU(2)$_R$ and singlets under SU(2)$_L$,
and the left-handed fermions behave reversely.
Thus a bidoublet Higgs field is required for the Yukawa couplings 
and also responsible for the electoweak symmetry breaking (EWSB).
The scale of the masses of the new gauge bosons in the LRM 
is constrained by direct searches and indirect analysis
\cite{czagon,cheung,chay,erler}, 
and we will discuss the constraints on the model in further detail.

This paper is organized as follows.
In section 2, we briefly review the charged sector in the general LRM. 
We explicitly show the right-handed current contributions
in the neutral $B$ meson system in section 3, 
and present the numerical analysis of $B^0-\bar{B}^0$ mixing 
and the dimuon charge asymmetry of $B$ mesons in the general LRM in section 4. 
Finally we conclude in section 5.

\section{The left-right model}

We briefly review the main features of the LRM, which are necessary for our analysis.
The gauge group of the left-right symmetric model is
SU(2)$_L \times$SU(2)$_R \times$U(1).
There exist a bidoublet Higgs field $\phi (2, \bar{2}, 0)$
and two triplet Higgs fields,
$\Delta_L (3,1,2)$ and $\Delta_R (1,3,2)$
in the minimal LR model
represented by
\be
\phi =
\left(
\begin{array}{cc}
\phi_1^0 & \phi_1^+ \\
\phi_2^- & \phi_2^0 \\
\end{array}
\right),
~~~~~~~~~~~~
\Delta_{L,R} =
\frac{1}{\sqrt{2}}
\left(
\begin{array}{cc}
\delta^+_{L,R} & \sqrt{2} \delta^{++}_{L,R}  \\
\sqrt{2} \delta^0_{L,R} & -\delta^{+}_{L,R}  \\
\end{array}
\right) ,
\ee
of which kinetic terms are given by
\be
{\cal L} = {\bf Tr} \left[ ( D_\mu \Delta_{L,R} )^\dagger
                       ( D^\mu \Delta_{L,R} )\right]
               + {\bf Tr} \left[ (D_\mu \phi)^\dagger
                        (D^\mu \phi) \right],
\ee
where the covariant derivatives are defined by
\bea
D_\mu \phi &=& \partial_\mu \phi - i \frac{g_L}{2} W^a_{L \mu} \tau^a \phi
              + i\frac{g_R}{2} \phi W^a_{R \mu} \tau^a ,
\nonumber\\
D_\mu \Delta_{L,R} &=& \partial_\mu \Delta_{L,R}
          - i\frac{g_{L,R}}{2} \left[W^a_{L,R \mu} \tau^a , \Delta_{L,R} \right]
          - i g^\prime B_\mu \Delta_{L,R}.
\eea
The gauge symmetries are spontaneously broken by
the vacuum expectation values (VEV)
\be
\langle \phi \rangle = \frac{1}{\sqrt{2}}\left(
\begin{array}{cc}
  k_1&0 \\
  0&k_2 \\
\end{array}
\right),
~~~~~~~~~~~~
\langle \Delta_{L,R} \rangle = \frac{1}{\sqrt{2}}\left(
\begin{array}{cc}
  0&0 \\
  v_{L,R}&0 \\
\end{array}
\right),
\ee
where $k_{1,2}$ are complex in general 
and $v_{L,R}$ are real, which lead to the charged gauge boson masses
\be
{\cal M}^2_{W^\pm} = \frac{1}{4} \left(
\begin{array}{cc}
g_L^2(k_+^2 + 2v_L^2) &  -2 g_L g_R k_1^\ast k_2 \\[3pt]
-2g_L g_R k_1 k_2^\ast   &  g_R^2 (k_+^2 +2v_R^2) \\
\end{array}
\right)
=
\left(
\begin{array}{cc}
M_{W_L}^2 & M_{W_{LR}}^2 e^{i\alpha} \\[3pt]
M_{W_{LR}}^2 e^{-i\alpha}  &  M_{W_R}^2 \\
\end{array}
\right) ,
\ee
where $k_+^2 = |k_1|^2 + |k_2|^2$ and $\alpha$ is the phase of $k_1^\ast k_2$.
Since the SU(2)$_{\rm R}$ breaking scale $v_R$
should be higher than the electroweak scale,
$k_{1,2} \ll v_R$, $W_R$ is heavier than $W_L$.
Note that $v_L$ is irrelevant for the symmetry breaking
and just introduced in order to manifest the left-right symmetry.
If the neutrino mass is purely determined by the see-saw relation
$m_\nu \sim v_{L}+k^2_{+}/v_{R}$,
$v_R$ is typically very large $\sim 10^{11}$ GeV.
It indicates that the heavy gauge bosons are too heavy
to be produced at the accelerator experiments
and the direct search of the SU(2)$_R$ structure is hardly achieved.
Therefore we assume that $v_R$ is only moderately large,
$v_R \sim {\cal O} ({\rm TeV})$,
for the heavy gauge bosons to be founnd at the LHC,
and the Yukawa couplings are suppressed
in order that the neutrino masses are at the eV scale.
We let $v_L$ be very small or close to 0 without loss of generality.
This is achieved when the quartic couplings of
$(\phi \phi \Delta_L \Delta_R)$-type terms in the Higgs potential
are set to be zero \cite{gunion2,kiers}
and warranted
by the approximate horizontal U(1) symmetry \cite{khasanov}
as well as the see-saw picture for light neutrino masses.
We adopt this limit here and note that the Higgs boson masses are
not affected by taking this limit \cite{kiers}.

The general Higgs potential in the LRM has been studied in Refs.
\cite{gunion1,gunion2,kiers}.
After the mass matrix is diagonalized by a unitary transformation, 
the mass eigenstates are written as
\be
\left(
\begin{array}{c}
W^\pm \\
W'^\pm \\
\end{array}
\right)
=
\left(
\begin{array}{cc}
\cos{\xi} & e^{-i\alpha}\sin{\xi} \\
-\sin{\xi} & e^{-i\alpha}\cos{\xi} \\
\end{array}
\right)
\left(
\begin{array}{c}
W_L^\pm \\
W_R^\pm \\
\end{array}
\right),
\ee
with the mixing angle
\be
\tan{2\xi} = -\frac{2 M_{W_{LR}}^2}{M_{W_R}^2 - M_{W_L}^2} .
\ee
For $v_R \gg |k_{1,2}|$, the mass eigenvalues and the mixing angle reduce to
\be
M^2_W \approx \frac{1}{4}g^2_L (|k_1|^2+|k_2|^2), \quad
M^2_{W^\prime} \approx \frac{1}{2}g^2_Rv_R^2, \quad
 \xi \approx \frac{g_L|k_1^*k_2|}{g_Rv_R^2}.
\ee
Here, the Schwarz inequality requires that 
$\zeta_g \equiv (g_R/g_L)^2\zeta \geq \xi_g \equiv (g_R/g_L)\xi$ 
where $\zeta \equiv M_W^2/M_{W'}^2 $.
From the global analysis of muon decay measurements 
\cite{MacDonald08}, 
the lower bound on $\zeta_g$ can be obtained without imposing discrete symmetry as follows:
\be
\zeta_g < 0.031 \qquad \textrm{or} \qquad M_{W^\prime} > (g_R/g_L) \times 460\
 \textrm{GeV} .
\label{MWRbound}
\ee
The new gauge boson mass $M_{W'}$ is severly constrained from $K_L - K_S$ mixing
if the model has manifest ($V^R = V^L$) left-right symmetry ($g_R = g_L$): 
$M_{W'} > 2.5$ TeV \cite{Zhang07},
where $V^L(V^R)$ is the left(right)-handed quark mixing matrice.
But, in general, the form of $V^R$ is not necessarily restricted 
to manifest or pseudomanifest ($V^R = V^{L*}K$) symmetric type,
where $K$ is a diagonal phase matrix \cite{LRM}.
Instead, if we take the following form of $V^R$, 
the limit on $M_{W'}$ may be significantly relaxed to approximately 300 GeV,
and the $W'$ boson contributions to $B_{d(s)}\bar{B}_{d(s)}$ mixings can be large 
\cite{Langacker89}:
\be
V^R_I = \left( \begin{array}{ccc} e^{i\omega} & \sim 0 & \sim 0 \\
               \sim 0 & c_R e^{i\alpha_1} & s_R e^{i\alpha_2} \\
         \sim 0 & -s_R e^{i\alpha_3} & c_R e^{i\alpha_4} \end{array} \right) ,\quad
V^R_{II} = \left( \begin{array}{ccc} \sim 0 & e^{i\omega} & \sim 0 \\
               c_R e^{i\alpha_1} & \sim 0 & s_R e^{i\alpha_2} \\
           -s_R e^{i\alpha_3} & \sim 0 & c_R e^{i\alpha_4} \end{array} \right) ,
\label{UR}
\ee
where $c_R\ (s_R)\equiv \cos\theta_R\ (\sin\theta_R)$ $(0^\circ \leq \theta_R \leq 90^\circ )$.
Here the matrix elements indicated $\sim 0$ may be $\lesssim 10^{-2}$ and the unitarity
requires $\alpha_1+\alpha_4=\alpha_2+\alpha_3$.
From the $b\rightarrow c$ semileptonic decays of the $B$ mesons, 
we can get an approximate bound $\xi_g\sin\theta_R \lesssim 0.013$ 
by assuming $|V^L_{cb}|\approx 0.04$ \cite{Voloshin97}.

\section{$B^0-\bar{B^0}$ mixing}

The neutral $B_q$ meson system $(q=d,s)$
is described by the Schr\"odinger equation
\be
i \frac{d}{dt} \left(
\begin{array}{c}
B_q(t) \\
\bar{B}_q(t) \\
\end{array}
\right)
= \left( M - \frac{i}{2} \Gamma \right)
\left(
\begin{array}{c}
B_q(t) \\
\bar{B}_q(t) \\
\end{array}
\right),
\ee
where $M$ is the mass matrix and $\Gamma$ the decay matrix. 
The $\Delta B = 2$ transition amplitudes
\be
\langle B_q^0 | {\cal H}_{\rm eff}^{\Delta B = 2} | \bar{B}_q^0 \rangle
         = M_{12}^q,
\label{M12}
\ee
yields the mass difference between the heavy and the light states of $B$ meson,
\be
\Delta M_q \equiv M_H^{q} - M_L^{q} = 2 | M_{12}^q |,
\ee
where $M_H^{q}$ and $M_L^{q}$
are the mass eigenvalues for the heavy and the light eigenstates, respectively.
The decay width difference is defined by
\be
\Delta \Gamma_q \equiv \Gamma_L^q - \Gamma_H^q
                = 2 |\Gamma_{12}^q| \cos \phi^q,
\ee
where the decay widths $\Gamma_L$ and $\Gamma_H$ are corresponding to
the physical eigenstates $B_L$ and $B_H$, respectively,
and the CP phase is
$\phi^q \equiv {\rm arg}\left( - M_{12}^q/\Gamma_{12}^q \right)$.
The charge asymmetry in Eq. (3) is expressed as
\be \label{aslq}
a_{sl}^q = \frac{|\Gamma_{12}^q|}{|M_{12}^q|} \sin \phi^q
         = \frac{\Delta \Gamma_q}{\Delta M_q} \tan \phi^q,
\ee
of which the SM predictions are given by
\cite{Lenz07}
\bea \label{aslSM}
a_{sl}^{d} = (-4.8^{+1.0}_{-1.2}) \times 10^{-4},
\quad
a_{sl}^{s} = (2.1 \pm 0.6) \times 10^{-5}, \cr
\phi^d = (-9.1^{+2.6}_{-3.8})\times 10^{-2},
\quad
\phi^s = (4.2\pm 1.4)\times 10^{-3} .
\eea
In the SM, $\Delta \Gamma_d/\Gamma_d$ is less than $1 \%$,
while $\Delta \Gamma_s/\Gamma_s \sim 10 \%$ is rather large.
The decay matrix elements $\Gamma_{12}^q$ is obtained from
the tree level decays $b \to c \bar{c} q$
where the dominent right-handed current contribution 
is suppressed by the heavy right-handed gauge boson mass $M_{W_R}$ \cite{Ecker86}.
Therefore, we ignore the contributions of our model to $\Gamma_{12}^q$ in this work.

We first consider the right-handed current contributions in the $B_d^0-\bar{B_d^0}$ system.
The $\Delta B = 2$ transition amplitudes in Eq. (\ref{M12}) is given by the following
effective Hamiltonian in the LRM \cite{Nam02}:
\be
H^{B\bar{B}}_{eff} = H^{SM}_{eff} + H^{RR}_{eff} + H^{LR}_{eff},
\ee
where
\bea H^{SM}_{eff} &=&
\frac{G_F^2M_W^2}{4\pi^2}(\lambda_t^{LL})^2S(x^2_t)
                (\bar{d_L}\gamma_\mu b_L)^2 , \label{HSMeff}\\
H^{LR}_{eff} &=& \frac{G_F^2M_W^2}{2\pi^2} \biggl\{
                  \left[\lambda_c^{LR} \lambda_t^{RL}x_cx_t\zeta_g A_1(x_t^2,\zeta)
                 + \lambda_t^{LR} \lambda_t^{RL}x_t^2\zeta_g A_2(x_t^2,\zeta)\right]
                   (\bar{d_L}b_R)(\bar{d_R}b_L) \cr
&&\hspace{0.2in} +\ \lambda_t^{LL} \lambda_t^{RL}x_b\xi_g^-
                    \left[x_t^3A_3(x_t^2)(\bar{d_L}\gamma_\mu b_L)(\bar{d_R}\gamma_\mu b_R)
                    + x_tA_4(x_t^2)(\bar{d_L}b_R)(\bar{d_R}b_L)\right]\biggr\} ,
\label{HLReff}
\eea
and
\be
\lambda_i^{AB} \equiv V_{id}^{A*}V_{ib}^{B} , \quad x_i \equiv \frac{m_i}{M_W} \ (i = u,c,t),
\quad \xi_g^{\pm} \equiv e^{\pm\alpha}\xi_g ,
\ee
with
\bea \label{loopfn}
S(x) &=& \frac{x(4 - 11x + x^2)}{4(1-x)^2} - \frac{3x^3\ln x}{2(1-x)^3} , \cr
A_1(x,\zeta) &=& \frac{(4-x)\ln x}{(1-x)(1-x\zeta)}
                 + \frac{(1-4\zeta)\ln \zeta}{(1-\zeta)(1-x\zeta)} , \cr
A_2(x,\zeta) &=& \frac{4-x}{(1-x)(1-x\zeta)} + \frac{(4-2x+x^2(1-3\zeta))\ln x}
                 {(1-x)^2(1-x\zeta)^2} + \frac{(1-4\zeta)\ln\zeta}{(1-\zeta)(1-x\zeta)^2} , \\
A_3(x) &=& \frac{7-x}{4(1-x)^2} + \frac{(2+x)\ln x}{2(1-x)^3} , \cr
A_4(x) &=& \frac{2x}{1-x} + \frac{x(1+x)\ln x}{(1-x)^2} . \nonumber
\eea
Note that $S(x)$ is the usual Inami-Lim function, 
$A_1(x,\zeta)$ is obtained by taking the limit
$x_c^2=0$, and $H_{eff}^{RR}$ is suppressed because it is proportional to $\zeta^2$.
Also in the case of $V^R_I$, one can see from Eq. (\ref{UR}) that there is no significant contribution of
$H^{LR}_{eff}$ to $B_d^0-\bar{B_d^0}$ mixing, 
so we only consider the $V^R_{II}$ type mixing matrix for $B_d^0-\bar{B_d^0}$ mixing.
The dispersive part of the $B_d^0-\bar{B_d^0}$ mixing matrix element can then be written as
\be \label{massmixing}
M_{12}^d = M_{12}^{SM} + M_{12}^{LR} = M_{12}^{SM}\left( 1 + r^d_{LR} \right) ,
\ee
where
\be
r^d_{LR} \equiv \frac{M_{12}^{LR}}{M_{12}^{SM}}
  = \frac{\langle \bar{B_d^0}|H_{eff}^{LR}|B_d^0 \rangle}
         {\langle \bar{B_d^0}|H_{eff}^{SM}|B_d^0 \rangle} .
\ee

For specific phenomenological estimates one needs the hadronic matrix elements of the operators
in Eqs. (\ref{HSMeff},\ref{HLReff}) in order to evaluate the mixing matrix element.
We use the following parametrization:
\bea
\langle \bar{B_d^0}|(\bar{d_L}\gamma_\mu b_L)^2|B_d^0 \rangle &=& \frac{1}{3}B_1f_B^2m_B , \cr
\langle \bar{B_d^0}|(\bar{d_L}\gamma_\mu b_L)(\bar{d_R}\gamma_\mu b_R)|B_d^0 \rangle
  &=& -\frac{5}{12}B_2f_B^2m_B , \\
\langle \bar{B_d^0}|(\bar{d_L}b_R)(\bar{d_R}b_L)|B_d^0 \rangle &=& \frac{7}{24}B_3f_B^2m_B \nonumber,
\eea
where
\be
\langle 0|\bar{d_\beta}\gamma^\mu\gamma_5 b_\alpha|B_d^0 \rangle 
= - \langle \bar{B_d^0}|\bar{d_\beta}\gamma^\mu\gamma_5 b_\alpha|0 \rangle
 = - \frac{if_Bp_B^\mu}{\sqrt{2m_B}}\frac{\delta_{\alpha\beta}}{3} ,
\ee
and where $f_B$ is the $B$ meson decay constant and $B_i\ (i=1,2,3)$ are the bag factors.
In the vacuum-insertion method \cite{Gaillard}, $B_i=1$ in the limit $m_b\simeq m_B$.
We will use $f_BB_i^{1/2}=(216\pm 15)$ MeV for our numerical estimates \cite{HPQCD09}.
Using the standard values of the quark masses and
$|V^L_{cd}|\approx 0.225$, one can express $r^d_{LR}$
in terms of the mixing angle and phases in the case of $V_{II}^R$ in Eq. (\ref{UR}) as
\bea \label{rLRd}
r^d_{LR} &\approx&  17.5 \biggl(
 \frac{1 - \zeta_g - (4.08 - 16.3\zeta_g)\ln(1/\zeta_g)) }{ 1 - 5.58\zeta_g }\biggr)
 \zeta_g s_R^2e^{-i(2\beta - \alpha_2 + \alpha_3)} \cr
 &-& 756 \biggl( \frac{ 1 - 5.03\zeta_g - (0.490 - 1.96\zeta_g )\ln(1/\zeta_g) }
         {1 - 10.2\zeta_g + 30.1\zeta_g^2}\biggr) \zeta_g s_Rc_Re^{-i(\beta + \alpha_3 - \alpha_4)} \
         - \ 7.94\xi_g s_R e^{-i(\beta + \alpha_3)} ,
\eea
where the mixing phase $\alpha$ was absorbed in $\alpha_i$ by redefining $\alpha_i + \alpha \rightarrow \alpha_i$,
and we used the approximation $A_i(x,\zeta) \simeq A_i(x,\zeta_g) (i=1,2)$
because $\zeta$ dependence on $A_i$ in Eq. (\ref{loopfn}) is rather weak for $M_{W'} > 100$ GeV
unless $g_R/g_L$ is drastically different from unity.

On the other hand, the right-handed current contributions to $B_s^0-\bar{B_s^0}$ mixing is sizable 
only in the case of $V_{I}^R$ as one can see from Eq. (\ref{UR}).
Similarly to $r^d_{LR}$, we obtain $r^s_{LR}$ in the case of $V_I^R$ as
\bea \label{rLRs}
r^s_{LR} &\approx&  -3.47 \biggl(
 \frac{1 - \zeta_g - (4.08 - 16.3\zeta_g)\ln(1/\zeta_g)) }{ 1 - 5.58\zeta_g }\biggr)
 \zeta_g s_R^2e^{-i(-\alpha_2 + \alpha_3)} \cr
 &+& 162 \biggl( \frac{ 1 - 5.03\zeta_g - (0.490 - 1.96\zeta_g )\ln(1/\zeta_g) }
         {1 - 10.2\zeta_g + 30.1\zeta_g^2}\biggr) \zeta_g s_Rc_Re^{-i(\alpha_3 - \alpha_4)} \
         + \ 1.70\xi_g s_R e^{-i\alpha_3}  .
\eea
The charge asymmtry $a^q_{sl}$ in Eq. (\ref{aslq}) can then be written in terms of $r^q_{LR}$ in the LRM as
\be
a^q_{LR}=a^q_{SM}\frac{\cos{\phi^q_{LR}}}{|1+r^q_{LR}|}
\left(1 + \frac{\tan{\phi^q_{LR}}}{\tan{\phi^q_{SM}}} \right), \quad
\phi^q_{LR} \equiv \textsl{arg}(1+r^q_{LR}),
\ee
where we omitted the subscript `$sl$'
and the SM values of $a^q_{sl}$ and $\phi^q$ are given in Eq. (\ref{aslSM}).
We use the above results for our numerical investigation of the right-handed current contributions
to the like-sign dimuon charge asymmetry in semi-leptonic $B$ decays in the next section.

\section{Results}

For our numerical analysis, 
we use the following $2\sigma$ bounds obtained from the deviation of
the present experimental data from the SM predictions on $B$ meson mixing \cite{Lenz11}:
\be \label{mixing_exp}
0.62 < |1+r^d_{LR}| < 1.15, \quad 0.79 < |1+r^s_{LR}| < 1.23 .
\ee
Note from Eqs. (\ref{rLRd},\ref{rLRs}) that we have six independent new parameters
($\zeta_g, \xi_g, \theta_R, \alpha_{2,3,4}$), and it is beyond the scope of this paper
to perform a complete analysis by varying all six parameters.
For simple illustration of the possible effect of the new interaction on $B$ meson mixing, instead,
we set $\xi_g = \zeta_g/2$ and $\alpha_{2,4}=0$
because $\xi_g$ contributions to $B$ meson mixing is expected to be much smaller than  $\zeta_g$'s
and  $\alpha_3$ is important as the overall phase of $r^q_{LR}$.

\begin{figure}[!hbt]
\centering%
    \includegraphics[width=8.3cm]{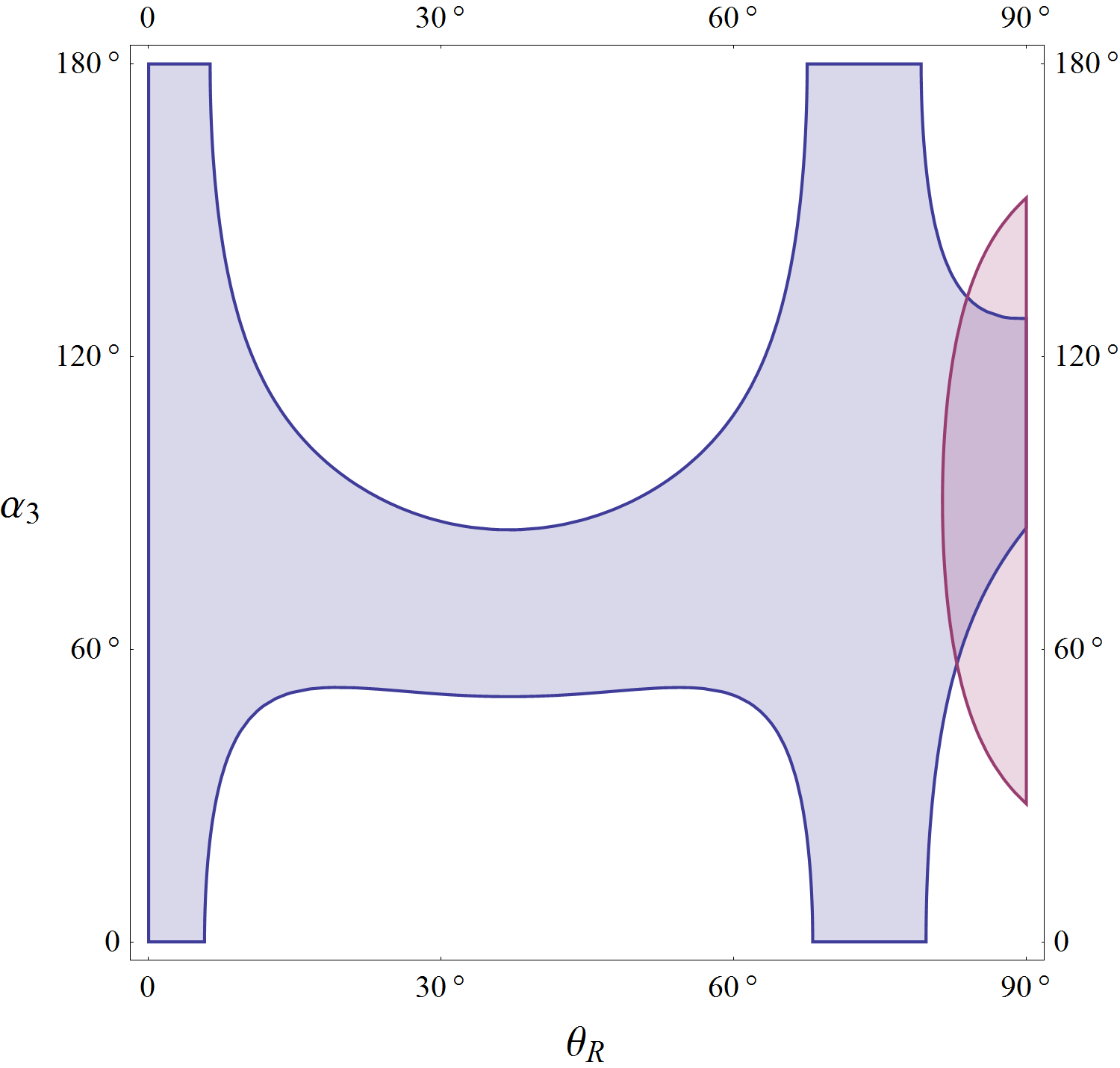}
\caption{Allowed regions for $\alpha_3$ and $\theta_R$ at 2 $\sigma$ level for $M_{W'} = 800$ GeV
in the case of $V_I^R$.
The red and blue regions are allowed by the current measurements
of the like-sign dimuon charge asymmetry and $B_s\bar{B_s}$ mixing, respectively.}
\label{BsMmix1}
\end{figure}

\begin{figure}[!hbt]
\centering%
    \includegraphics[width=8cm]{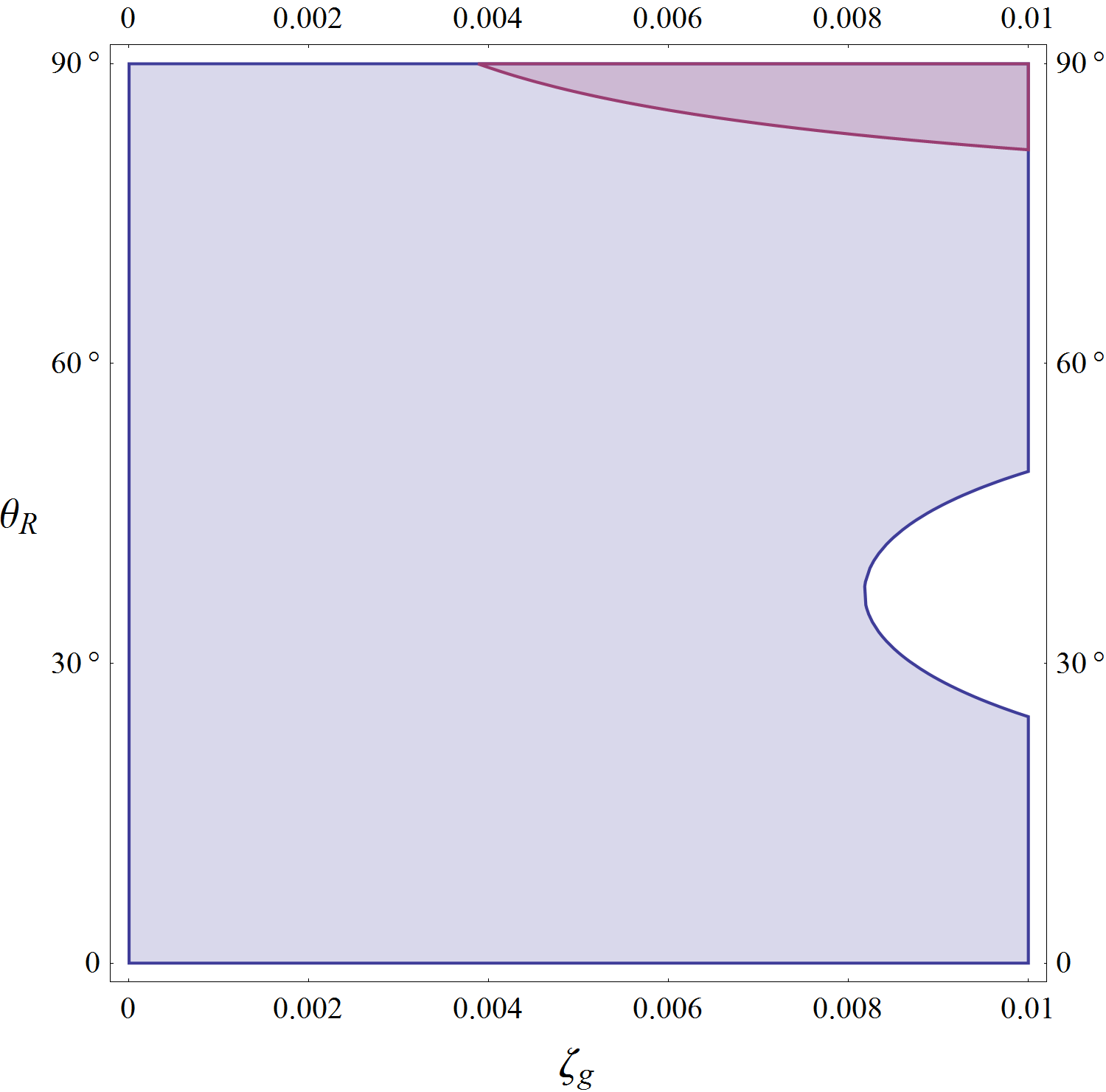}
\caption{Allowed regions for $\theta_R$ and $\zeta_g$ at 2 $\sigma$ level for $\alpha_3=90^\circ$
in the case of $V_I^R$.
The red and blue regions are allowed by the current measurements
of the like-sign dimuon charge asymmetry and $B_s\bar{B_s}$ mixing, respectively.}
\label{BsMmix2}
\end{figure}

In the case of $V_I^R$, as discussed earlier, 
the right-handed current contributions to $B_s-\bar{B_s}$ mixing could be sizable
while those to $B_d-\bar{B_d}$ mixing is negligible.
With the present experimental bounds of the dimuon charge asymmetry 
and $B_s-\bar{B_s}$ mixing given in Eqs. (\ref{asym_exp},\ref{mixing_exp}), 
we first plot the allowed region of $\alpha_3$ and $\theta_R$
for $M_{W'} = 800$ GeV at 2 $\sigma$ level in Fig. \ref{BsMmix1}.
One can see from the overlapped allowed region in the figure 
that large values of $\theta_R$ are preferred.
This is the clear indication that manifest or pseudomanifest LRM is disfavored in this case.
In Fig. \ref{BsMmix2}, 
we plot the allowed region of $\theta_R$ and $\zeta_g$ for $\alpha_3=90^\circ$
at 2 $\sigma$ level.  One can obtain the lower bound of $\zeta_g \gtrsim 0.004$ from the figure
which corresponds to the upper bound of $W'$ mass $M_{W'} \lesssim (g_R/g_L)\times 1.3$ TeV.
For different values of $\alpha_3$, this mass bound can change, but not very much.
In other words, 
if it happens that the mass of $W'$ is much larger than the obtained upper bound,
the right-handed contributions are not big enough 
to explain the present measurement of the dimuon charge asymmetry.

\begin{figure}[!hbt]
\centering%
    \includegraphics[width=8.3cm]{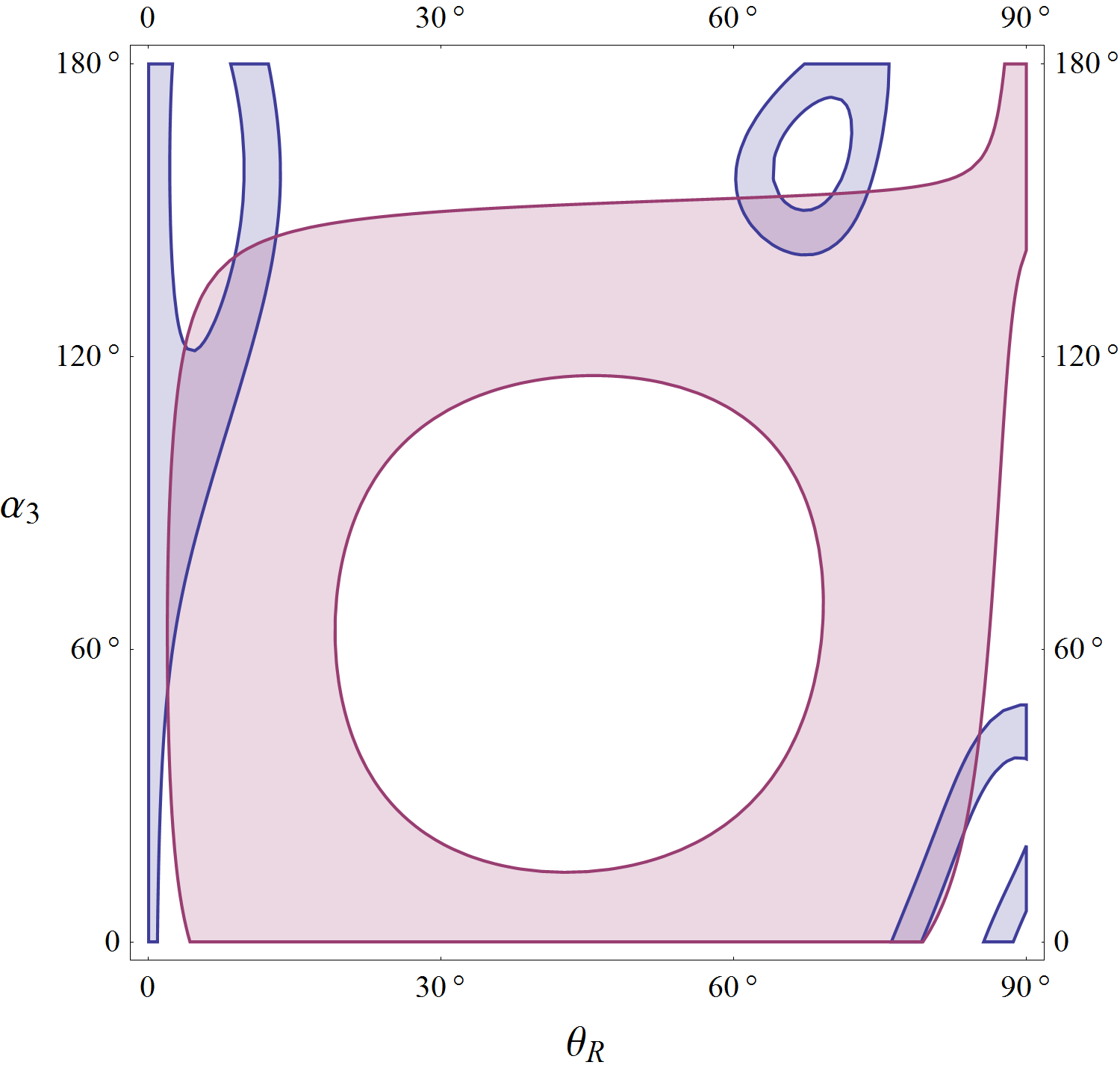}
\caption{Allowed regions for $\alpha_3$ and $\theta_R$ at 2 $\sigma$ level for $M_{W'} = 800$ GeV
in the case of $V_{II}^R$.
The red and blue regions are allowed by the current measurements
of the like-sign dimuon charge asymmetry and $B_d\bar{B_d}$ mixing, respectively.}
\label{BdMmix1}
\end{figure}

\begin{figure}[!hbt]
\centering%
    \includegraphics[width=8cm]{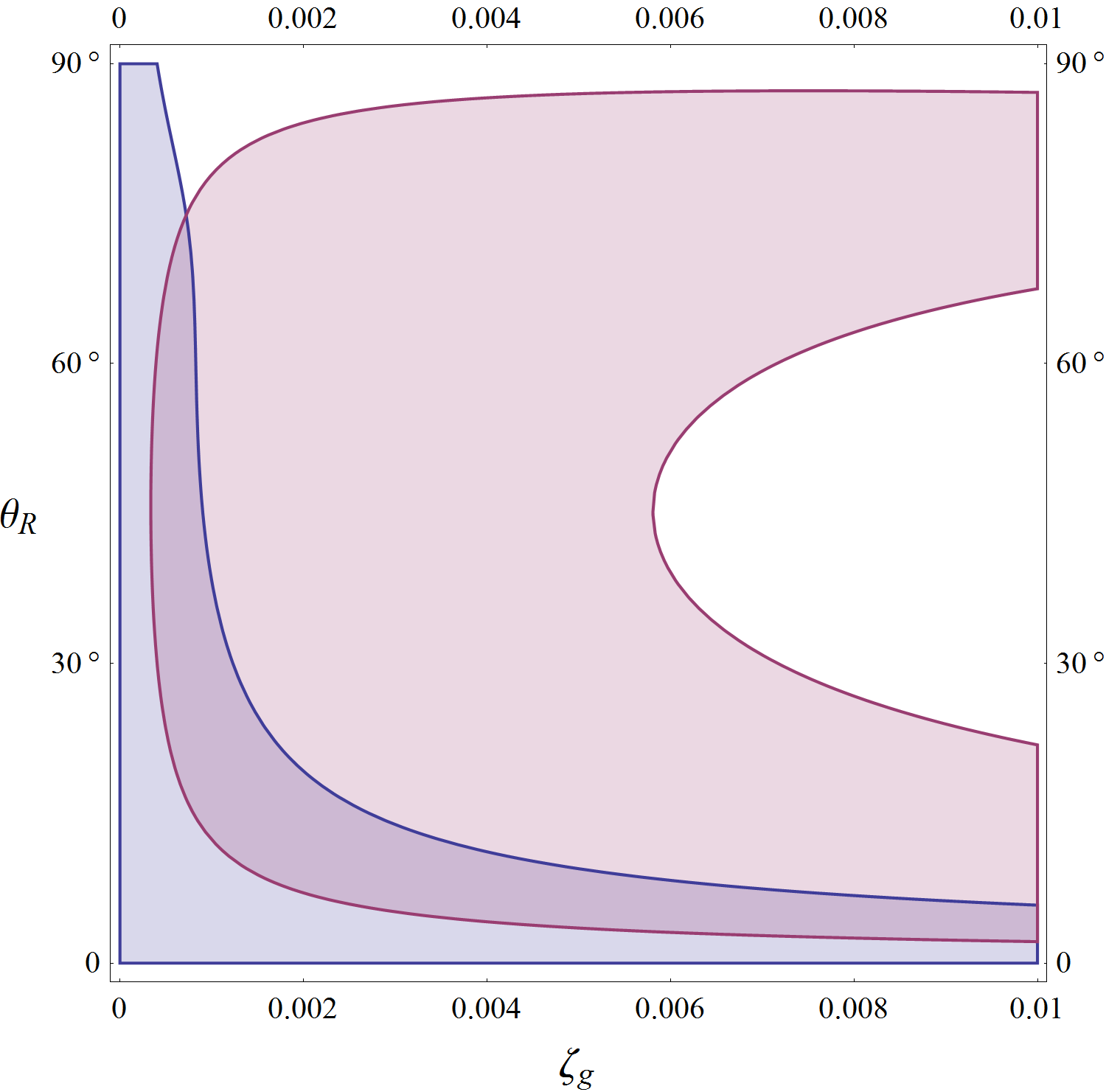}
\caption{Allowed regions for $\theta_R$ and $\zeta_g$ at 2 $\sigma$ level for $\alpha_3=90^\circ$
in the case of $V_{II}^R$.
The red and blue regions are allowed by the current measurements
of the like-sign dimuon charge asymmetry and $B_d\bar{B_d}$ mixing, respectively.}
\label{BdMmix2}
\end{figure}

In the case of $V_{II}^R$, on the other hand, 
the right-handed current contributions to $B_d-\bar{B_d}$ mixing could be sizable
while those to $B_s-\bar{B_s}$ mixing is negligible.
Similarly to the $V_I^R$ case, we plot the allowed region of  $\alpha_3$ and $\theta_R$
for $M_{W'} = 800$ GeV at 2 $\sigma$ level in Fig. \ref{BdMmix1}.
The figure shows that small or large values of $\theta_R$ are allowed unlike the $V_I^R$ case.
In order for direct comparison with the $V_I^R$ case,
we plot again the allowed region of $\theta_R$ and $\zeta_g$ for $\alpha_3=90^\circ$
at 2 $\sigma$ level in Fig. \ref{BdMmix2}.  The figure shows that $V_{II}^R$ senario allows more wide range of
allowed area of new parameter space and the lower bound of $\zeta_g$ is approximately $\zeta_g \gtrsim 0.0004$.
We obtain the corresponding upper bound of $W'$ mass $M_{W'} \lesssim (g_R/g_L)\times 4$ TeV.
We found that this mass bound could be somewhat lower for different values of $\alpha_3$.
It should also be noted that we have similar results for different $\alpha_{2,4}$ in both senarios.

\section{Concluding Remarks}
In this paper, 
we studied the right-handed current contributions to 
the CP violating like-sign dimuon charge asymmetry
in semi-letonic $B$ decays in general left-right models.
Without imposing manifest or pseudomanifest left-right symmetry, 
we consider two types of mass mixing matrix $V^R$
with which $W'$ contributions are big enough to explain the current mismatch of
the present measurents of the dimuon charge asymmetry and the SM prediction.
We evaluated the sizes of $W'$ contributions to $B_d-\bar{B_d}$ and $B_s-\bar{B_s}$ mixings
which govern the dimuon charge asymmetry, 
and obtained the allowed regions of NP parameter spaces.
With the given parameter sets, we have the following mass bounds of $W'$:
$M_{W'} \lesssim (g_R/g_L)\times 1.3$ TeV for Type I ($V_I^R$) or
$M_{W'} \lesssim (g_R/g_L)\times 4$ TeV for Type II ($V_{II}^R$),
which represent the amount of NP effects enough to explain the present measurent of the dimuon charge asymmetry.
If we consider the early LHC bound on $W'$ \cite{LHCW},
Type I model including manifest or pseudomanifest LRM is disfavored if $g_R = g_L$.
This analysis can affect other $B$ meson mixing related observables such as $\sin{2\beta}$
and mixing induced CP violation in B decays.
A detailed discussion on such mixing induced CP asymmetries in general LRM can be found in Ref. \cite{Nam03},
and a combined study including other decays with new experimental results will be discussed in the follow-up paper.

\acknowledgments
KYL is supported in part by WCU program through the KOSEF funded by the MEST (R31-2008-000-10057-0)
and the Basic Science Research Program through the National Research Foundation of Korea (NRF)
funded by the Korean Ministry of Education, Science and Technology (2010-0010916).

\def\PRD #1 #2 #3 {Phys. Rev. D {\bf#1},\ #2 (#3)}
\def\PRL #1 #2 #3 {Phys. Rev. Lett. {\bf#1},\ #2 (#3)}
\def\PLB #1 #2 #3 {Phys. Lett. B {\bf#1},\ #2 (#3)}
\def\NPB #1 #2 #3 {Nucl. Phys. {\bf B#1},\ #2 (#3)}
\def\ZPC #1 #2 #3 {Z. Phys. C {\bf#1},\ #2 (#3)}
\def\EPJ #1 #2 #3 {Euro. Phys. J. C {\bf#1},\ #2 (#3)}
\def\JHEP #1 #2 #3 {JHEP {\bf#1},\ #2 (#3)}
\def\IJMP #1 #2 #3 {Int. J. Mod. Phys. A {\bf#1},\ #2 (#3)}
\def\MPL #1 #2 #3 {Mod. Phys. Lett. A {\bf#1},\ #2 (#3)}
\def\PTP #1 #2 #3 {Prog. Theor. Phys. {\bf#1},\ #2 (#3)}
\def\PR #1 #2 #3 {Phys. Rep. {\bf#1},\ #2 (#3)}
\def\RMP #1 #2 #3 {Rev. Mod. Phys. {\bf#1},\ #2 (#3)}
\def\PRold #1 #2 #3 {Phys. Rev. {\bf#1},\ #2 (#3)}
\def\IBID #1 #2 #3 {{\it ibid.} {\bf#1},\ #2 (#3)}

\end{document}